\documentclass[aps,prl,reprint,superscriptaddress,showpacs,longbibliography]{revtex4-1}
\usepackage{graphicx}
\usepackage{hyperref}
\usepackage{epstopdf}
\usepackage{xcolor}
\newcommand{\cred}[1]{{\color{black} #1}}

\begin{document}
\title{Thermally driven inhibition of superconducting vortex avalanches}
\author{Antonio Lara}
\author{Farkhad G. Aliev}
\thanks{farkhad.aliev@uam.es}
\affiliation{Dpto. F\'isica de la Materia Condensada, INC and IFIMAC, Universidad Aut\'onoma de Madrid. 28049 Madrid, Spain}
\author{Victor V. Moshchalkov}
\affiliation{INPAC-Katholieke Universiteit Leuven, Celestijnenlaan 200D, B3001, Leuven,
Belgium}
\author{Yuri~M.~Galperin}
\affiliation{Department of Physics, University of Oslo, 0316 Oslo, Norway}
\affiliation{A.~F.~Ioffe Physico-Technical Institute of Russian
Academy of Sciences, 194021 St. Petersburg, Russia}

\begin{abstract}
Complex systems close to their critical state can exhibit abrupt transitions -- avalanches -- between their metastable states. It is a challenging task to understand the mechanism of the avalanches and control their behavior.
Here we investigate microwave ($mw$) stimulation of avalanches in the so-called vortex matter of type II superconductors - a system of interacting Abrikosov vortices close to the critical (Bean) state. 
Our main finding  is that the avalanche incubation strongly depends on the excitation frequency, a completely unexpected behavior observed close to the so-called depinning frequencies. Namely, the triggered vortex avalanches in Pb superconducting films become effectively inhibited approaching the critical temperature or critical magnetic field when the $mw$ stimulus is close to the vortex depinning frequency. We suggest a simple model explaining the observed counter-intuitive behaviors as a manifestation of the strongly nonlinear dependence of the driven vortex core size on the $mw$ excitation intensity. This paves the way to controlling avalanches in 
\cred{superconductor-based devices}
%complex systems 
through their nonlinear response.

\end{abstract}
\maketitle

\section{Introduction}
\cred{Control over stability of vortex matter in type II superconducting films and materials is essential for broad range of current and future potential applications. Penetration of magnetic flux in form of vortex avalanches is the most harmful for the real devices ranging from microwave resonators, bolometers, superconducting quantum interference devices or superconducting tapes (see, e.g.,~\cite{chigo}.
}

In general, an avalanche is an apparently unpredictable abrupt transition between two metastable states.
Snow avalanches \cite{avalanch_hotspot}, sand avalanches and woodquakes~\cite{woodquakes_PRL2015} are some widely known examples of  such events  in natural systems.  Avalanches in interacting networks of strongly nonlinear dynamical nodes are currently considered the  main fundamental mechanisms describing biological information processing~\cite{Stopp2016,Beggs2003}. 

The avalanches widely explored in physics and technology~\cite{avalanch_photodiode,Bulgarini2012,Jukimenko2014},
can interfere in the operation of many devices. An example is magnetic avalanches
in superconducting resonators~\cite{chigo}, implemented in a wide range of systems from quantum bits to devices for research in astrophysics. These avalanches manifest themselves  as 
spontaneous penetrations of magnetic flux in the bulk of the superconductor~ \cite{Field95,Johansen2002_EPL}.
\cred{
The number of applications based on superconductors increases, and with the important role
these devices will play in technology~\cite{Wang}, safe operation becomes crucially important.}

According to the model based on thermomagnetic instabilities~\cite{Rakhmanov2004}, an avalanche is triggered by a thermal fluctuation (hot spot), which facilitates more flux motion toward the hot place, with a subsequent heat release. Under quasi-equilibrium conditions, the avalanches can also be triggered by microwave pulses~\cite{johansen2002,tejada} or AC signals at near resonant frequencies of superconducting cavities~\cite{chigo}. The situation is somewhat different with a broadband microwave ($mw$) field  sweep, with avalanches triggered at different depinning frequencies due complicated local vortex pinning potentials~\cite{awad2011}.

It is a great challenge to control avalanche processes~\cite{Zahar2013,Curcio2015}. In superconductors, the most common ways to control the fast flux instabilities were through nano-morphology~\cite{Yurchenko2013} or optimization of the heat transfer through coating a superconducting film with a metal layer~\cite{Choi2008}. Here we propose a means of avalanche control using the intrinsic nonlinear $mw$ response of the fundamental units forming the network, superconductor vortices. Such nonlinearity has been recently identified through microwave-stimulated superconductivity due to presence of vortices~\cite{lara2015}.

We report a new, completely unexpected effect: it turns out  that application of microwaves   at frequencies  close to the \textit{ vortex depinning} frequencies \textit{inhibits} avalanches. We attribute this effect to a nonlinear dependence of the core size of a dynamically driven vortex  
on its velocity.  This effect allows one to engineer the avalanche behavior, at least to some extent.

\section{Experimental details}
We studied thin films of Pb with a thicknesses of 60 and 70 nm. This thicknesses are well below 
the critical value  ensuring type II superconductivity ~\cite{Dolan}. In what follows we present data only from the 70 nm film, since the results observed in both samples are similar. Details in the samples growth have been reported before \cite{lara2015}.
The measurements have been done in a variable temperature JANIS helium cryostat with a superconducting magnet to apply static magnetic fields. The microwave signal is supplied by a gold coplanar waveguide (CPW),
 Fig. \ref{fig:1}a.  The microwave signal is generated and detected by a network analyzer, that carries it to the CPW. More details about this system and the measurement procedure can be found in~\cite{lara2015}. 
The presence of avalanches is detected as jumps in the normalized transmission parameter $S_{21}$ (we denote this normalized complex quantity as $U=U'+iU''$), measured by the network analyzer~\cite{chigo,awad2011}. 

\begin{figure}[h]
\includegraphics[width=0.7\linewidth]{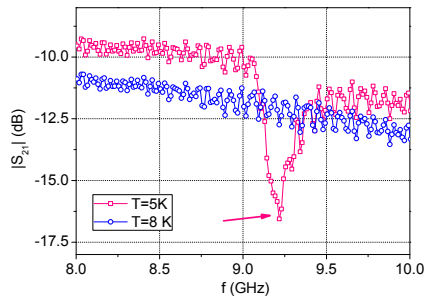}
\caption{Arrows point at depinning frequencies, where a large amount of energy is absorbed by the sample in the superconducting state with respect to the normal state}
\label{fig:2S}
\end{figure}

In this work we focus on the dependence of the microwave power required to trigger avalanches as a function of $T$ and $H$. 
The microwave frequency has been fixed either to values where the transmission parameter follows a flat tendency both in the superconducting and normal phases, or to specific values in which the superconducting phase exhibits a strong dip  in the transmission, at the so-called vortex depinning frequencies  $f_{DP}$~\cite{awad2011} (figure \ref{fig:2S}).
The temperature and magnetic field sweeps were extended up to the values destroying the superconductivity (above $T_c$ or $H_{c2}$), to use them as a reference.

We have also performed computer simulations of the Time-Dependent Ginzburg-Landau equation (TDGL) to propose a model to explain the unusual avalanche behavior with the interplay of vortex displacement and radius under a high frequency field. This program is described in more detail in~\cite{lara2015}. The results of these simulations are presented with the frequency in units of $f_0=96k_BT_c/\pi \hbar$ 
and field (including the high frequency $mw$ magnetic field) in units of $H_{c2}$.

\section{Results}
\begin{figure}[ht]
\includegraphics[width=\linewidth]{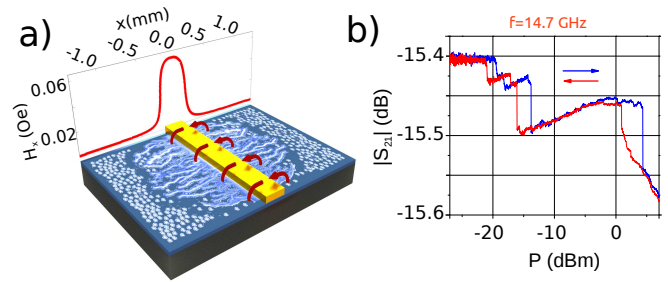}
\caption{a) Sketch of the sample.  The graph shows the calculated in plane magnetic field of the stripline.
The red arrows represent the $mw$ field. In light blue is a supposed representation of vortices and avalanches, generated mainly by the stripline. b) Module of microwave transmission $S_{21}$ vs microwave power, swept from minimum to maximum (blue) or viceversa (red) at a rate of 1 dBm/s, T=5 K.}
\label{fig:1}
\end{figure}

Figure~\ref{fig:1}a sketches the experiment. The superconducting film (blue), grown on top of a substrate (dark gray) is in contact with the microwave source (the coplanar waveguide, sketched in yellow). The latter generates a microwave magnetic field (red arrows), incident on the sample, whose parallel to the plane component is shown in the graph in Fig,~\ref{fig:1}a. Most of the $mw$ is concentrated close to the stripline, therefore the high frequency field exciting the sample is local.  This suggests that the avalanches would most probably be generated near the stripline (light blue lines). 

Our previous study demonstrated the presence of a set of vortex depinning frequencies in type II superconducting films~\cite{awad2011}. The present experiments aim on the study of these jumps as a function of applied microwave power, for a set of frequencies. The applied field is the remanent one after saturating at 1000~Oe and going back to zero field. In Fig.~\ref{fig:1}b, an example of these jumps is presented for a 70 nm thick Pb film. Depending on the direction of the power sweep (from the minimum to the maximum value, blue line, or viceversa, red) the jumps appear at different powers. This indicates that the release of an avalanche is influenced by the ``history" of the previously released ones. 
However, at the lowest (or highest) microwave powers, $|S_{21}|$ is at the same value, indicating that the global superconducting state is equivalent after all these avalanche 
events. Thus, a ``hysteretic" behavior in microwave power evidences the importance of thermal effects in avalanches. Additionally they show clearly the 
reproducibility of the power dependent transmission, so that the jumps in microwave permeability are fully controlled by the applied frequency and $mw$ power. We have not observed substantial changes in the 
shape of power dependent permeability on the sweep rate varied between 1 and 10 dBm/s. 

\begin{figure}[t]
\includegraphics[width=\linewidth]{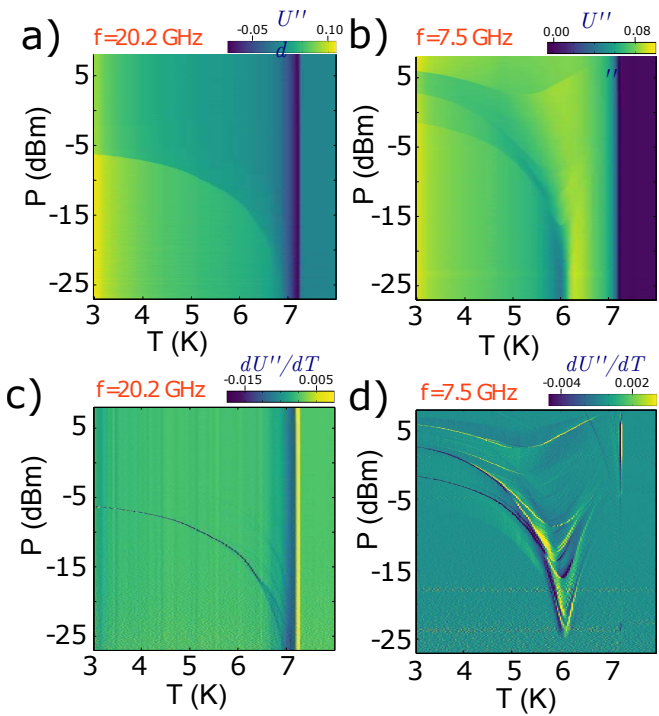}
\caption{Microwave stimulated avalanches in a 70 nm film at a freezing field of 10 Oe. Part a) shows $U''$ with expected avalanche behavior vs. temperature and power measured at $f=20.2$ GHz. Part (b) shows $U''$ vs. microwave power and temperature at different microwave frequency of $f=7.5$ GHz. Unexpected avalanche behavior vs. temperature, close to a depinning frequency is seen. Parts c) and d) are derivatives with respect to temperature of the parts a) and b). Color bars show  $U''$ and it´s derivatives in arbitrary units.}
\label{fig:2}
\end{figure}

Our main findings are presented in Fig.~\ref{fig:2}a)-d) where we compare a power and temperature sweep of the normalized transmission parameter at 20.2 GHz (a,c) and 7.5 GHz (b,d). One notices the previously mentioned (Fig.~\ref{fig:1}b) jumps as steps in the 2D color plot of microwave permeability evident in Fig.~\ref{fig:2}a. Moreover one clearly observes that for temperatures approaching $T_c$ (7.2~K) this jump requires less $mw$ power to occur. That is in agreement with the expected thermally driven avalanche behavior, since higher temperatures favor the entrance of magnetic flux and enhance the vortex mobility. The corresponding differential  plot (derivative with respect to temperature) shown in  Fig.~\ref{fig:2}c) helps to observe both the main avalanche (splitting into several approaching $T_c$). All the avalanches observed outside the vicinities of the 
depinning frequencies   show similar temperature dependences.  
The critical $mw$ power for triggering an avalanche monotonously \textit{decreases} while approaching $T_c$.

Figure~\ref{fig:2}b shows the same measurements as Fig.~\ref{fig:2}a, but close to one of the vortex depinning frequencies, $f_{DP}= 7.5$~GHz. One clearly observes an unexpected temperature dependence of the critical  $mw$ power. Approaching $T_c$, within some temperature range (around 6 K for the lowest powers, and around 5K for the strongest ones) the expected tendency of decreasing critical power inverts. The closer one approaches $T_c$, the more power is required to trigger the avalanches. We shall  further refer to this effect as thermally driven avalanche inhibition (TDAI). 

\begin{figure}[h]
\includegraphics[width=0.7\linewidth]{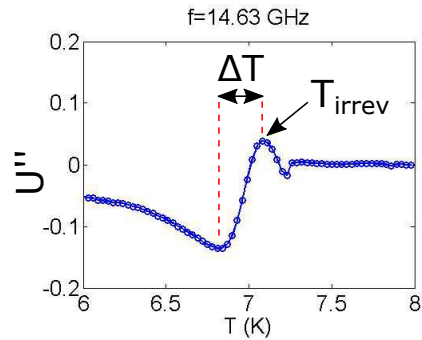}
\caption{Temperature dependence of the imaginary contribution to the losses $U''$ in microwave driven vortex system. The irreversibility temperature is defined as maximum in losses. Also shown in a temperature interval between maximum and minimum in losses which is described with error bars in Fig.~\ref{fig:3}e.}
\label{fig:1S}
\end{figure}

There is no simple relationship between the $mw$ frequency and presence or absence of the TDAI.
A careful examination of the power-temperature and power-field sweeps 
shows that each time the inversion of the normal tendency to the anomalous one happens, it takes place in a range of temperatures  where there is a visible maximum in microwave losses $U^{\prime \prime}$ (see Fig.~\ref{fig:1S}). 
The peak in $mw$ losses close to the $T_c$ is known to originate from the enhanced vortex mobility \cite{Coffey1992,lara2015}. 

\begin{figure}[t]
\includegraphics[width=\linewidth]{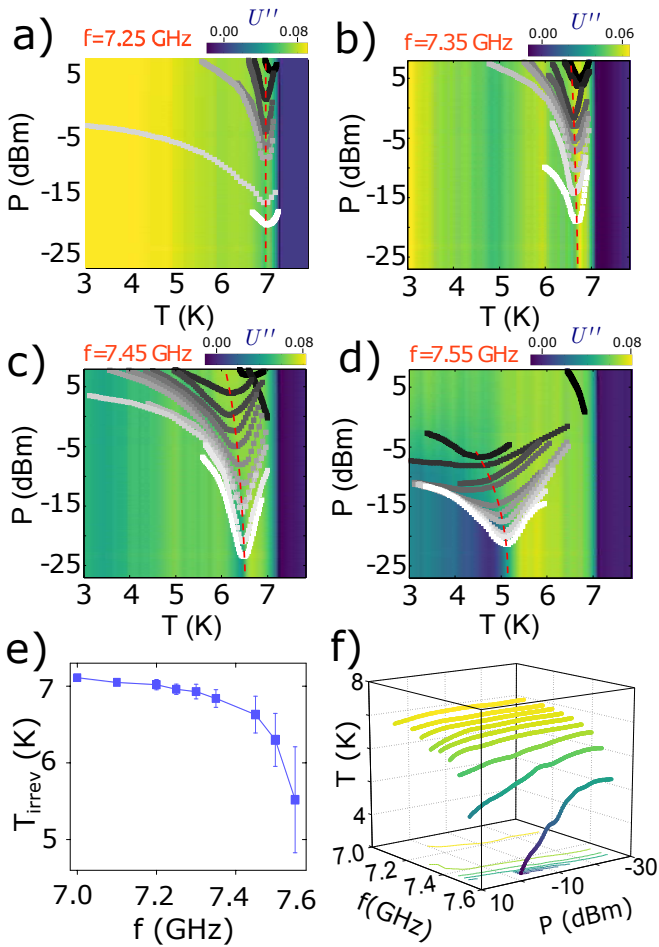}
\caption{a-d) Temperature dependence of microwave susceptibility close to one of the depinning frequencies at freezing field. Color bars show  $U''$ in arbitrary units. Different avalanche branches are indicated in gray tones, extracted from a differential analysis. e) Temperature of the $U''$ peak vs. frequency near a depinning frequency, for $P=-20$ dBm. Error bars represent the temperature difference between the maximum and the base of the peak  (see Figure \ref{fig:1S}). f) Dependence of the peak on temperature, power and frequency. }
\label{fig:3}
\end{figure}

Figure~\ref{fig:3} shows in more detail how the TDAIs vary in the proximity of one of vortex depinning frequencies $f_{DP}$.  At $f<f_{DP}$ the TDAI effects occur close to $T_c$ in the temperature range close to the peak in microwave losses $U^{\prime \prime}$ when the superconducting gap becomes partially filled by quasiparticles  and the vortices are mobile. A further increase of the drive frequency near $f=f_{DP}$ shifts the range where the TDAI initiate ing its temperature. Moreover, as could be observed independently of specific depinning frequencies analyzed in Figs.~\ref{fig:2}d and  \ref{fig:3}, the higher the applied powers is (i.e., less external temperature is needed to create the same vortex mobility), the lower is the temperature at which the corresponding inversion point in the temperature dependence of avalanches occurs. Figure~\ref{fig:3}e shows that the peak in the temperature dependence of $U^{\prime \prime}$ also moves towards lower temperatures as the $mw$ frequency approaches and exceeds the depinning frequency.

At yet higher frequencies, $f>f_{DP}$, the TDAI effects disappear (not shown), as the peak in  $U^{\prime \prime}$ vanishes on the lower temperature side of the scale. The scenario repeats again approaching a different vortex depinning frequency. The zero slope point at lower temperatures of the peak in losses shown in Fig.~\ref{fig:3}f) represents equal vortex mobility surface as a function of temperature, power and frequency close to particular vortex depinning frequency.

We have already mentioned that the TDAI may be expected to take place mainly close to the microwave source (the CPW central conductor). A few avalanches could, however, be triggered  far from  the $mw$ source. Those avalanches should then be subject to a reduced $mw$ power and be thermally stimulated mainly so that their temperature dependence could not invert approaching the critical temperature or critical field  (see
 Fig.~\ref{fig:3} and further below).

\begin{figure}
\includegraphics[width=\linewidth]{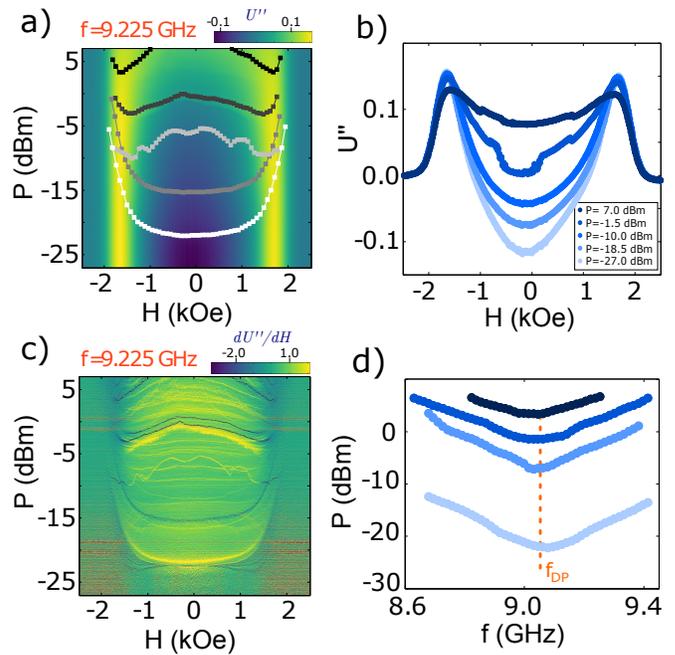}
\caption{a) Microwave losses  $U^{\prime \prime}$ as a function of applied field and microwave power measured at $T=0.7T_c$ ($T=5$~K) near depinning frequency $f_{DP}=9.225$~GHz. b) Cross sections of a) at different powers, highlighting the peaks in losses near the critical field. c) Differential plot ($dU^{\prime \prime}/dH$) of a). It allows to more clearly detect avalanches, marked with dots in a).  
d) Different frequency dependent avalanches close to the depinning frequency of a).}
\label{fig:4}
\end{figure}

Figure~\ref{fig:4}a presents a magnetic field driven analog of the TDAI effect, but now studied at a fixed temperature $T=5$~K and different $mw$ powers at some higher depinning frequency $f_{DP}=9.225$~GHz.
Figure~\ref{fig:4}c shows a differential plot ($dU^{\prime \prime}/dH$)  helping to resolve more clearly the main avalanches marked with dots in a). In the field range where the peaks in $U^{\prime \prime}$ are observed close to the upper critical field (see Fig.~\ref{fig:4}b), the avalanches show a change in tendency, 
requiring more power to be triggered at higher fields, where vortex mobility gets enhanced. Compared to temperature sweeps, field sweeps look more complex, with more avalanche branches present, and some of them suffering ``jumps" in field as well. This should not be surprising, as changing the applied field changes the number of vortices present in the sample and enhances non-equilibrium, which stimulates the avalanche process.

Figure~\ref{fig:4}d shows how critical power needed to stimulate avalanches varies as a function of frequency close to some specific $f_{DP}=9.225$ GHz.
\cred{At a fixed temperature ($T/T_c =0.7$) and some small magnetic field ($H\simeq100$ Oe) the $mw$ power needed to trigger avalanches reduces nearly symmetrically with frequency relative to the depinning frequency.}

\section{Model and discussion}
The experimentally observed unforeseen thermally induced enhanced stability of the vortex avalanches to the external microwave excitation could be understood as a consequence of a counter-intuitive dependence of the vortex core size on the  $mw$ power and frequency. 

Larkin and Ovchinnikov (LO)  were the first to point out a strongly nonlinear dependence on the vortex core size on the vortex 
velocity~\cite{LO1976}.
\begin{figure}[ht]
\includegraphics[width=\linewidth]{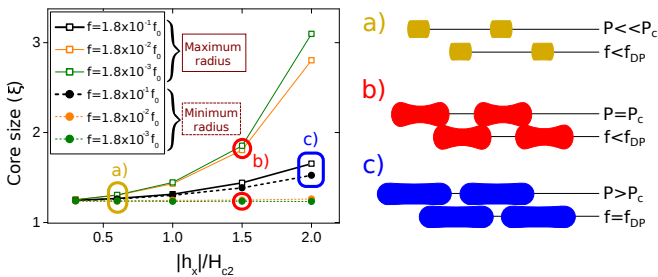}
\caption{Simulations of the change in vortex core size during an oscillation period for different frequencies and amplitudes of $rf$ field. The values of radius are taken at $\vert\Psi\vert=0.5$. The sketch at the right illustrates the overlap of the regions where vortices move under microwave fields of different powers. As vortices move, their radius change. This could allow heated zones to overlap to produce avalanches. The situations a), b) and c) correspond those indicated in the graph with the same letters and colors.}
\label{fig:5}
\end{figure}
They predicted a reduction of the vortex core size as the vortex velocity approaches some critical value.
As a consequence of the LO effect, the size of the core of a periodically driven vortex changes in time.
In order to understand the experimental results we consider for simplicity three qualitatively different regimes. In the regimes a) with low or b) with high $mw$ power (microwave magnetic field) but outside the vortex depinning regime, the LO effect  manifests itself only in the regions of  maximal velocity (i.e., close to displacement minima). On the other hand, in the regime c), which is  close to a depinning frequency, the  vortices become more mobile. As a result,  the LO reduction of vortex core  manifests itself during the whole displacement cycle.

To elucidate the physics of the thermally inhibited avalanches in the driven  vortex system, following 
the approach~\cite{lara2015}, we have simulated the relative changes in the vortex core size with both drive frequency and microwave power.  Figure~\ref{fig:5} explains the model describing a transition from thermally activated to thermally suppressed vortex avalanches approaching a depinning frequency. Based on the simulation results, we conclude that for the driving frequencies below the depinning frequency and low $mw$ power $P$  local ``normal'' regions created by the driven vortex do not overlap and are not sufficient to trigger the avalanche process (case a) remarked in yellow).  Note that the depinning frequency $f_{DP}$ plays the role of a critical value above which the core size shrinks under $mw$ drive~\cite{lara2015}.
Increasing the $mw$ power, but still below the depinning frequency, enhances the absolute changes in the vortex core size (Fig.~\ref{fig:5}), inducing the overlap between different extended core areas and triggering the avalanche close to some critical power $P=P_c$ (case b) in simulations marked in red). Evidently, such process is thermally activated, as long as the vortex diameter strongly increases approaching critical temperature $T_c$ .

However, close to depinning frequencies and in the temperature region where vortices become depinned (i.e., at temperatures about $T_{DP}$), the LO effect reduces the vortex core ending up in only small variation in the vortex core size during periodically driven motion (case c) Fig.~\ref{fig:5}). This reduces the probability of the overlap between normal regions and, therefore, the avalanche should be triggered at yet higher applied $mw$ powers for the fixed temperature, exactly as we observe experimentally.

The proposed scenario suggests only the basic mechanism behind the TDAI effect. Specific forms of the normal areas occupied by the $mw$ driven vortex should depend on the pinning details and thermal energy relaxation rate outside the vortex core. However, once the percolation on the local level  between the ``hot spots'' created by the driven vortices is established, the avalanche could be triggered with high probability. Superconducting vortex avalanches are usually described by a thermomagnetic instability. In this model, the avalanches develop as a result of the heat released due to  flux  motion leading to enhancement of the local vortex mobility and correspondingly further heat release. This positive feedback balanced by heat dissipation towards the substrate results in ultrafast dendritic flux redistributions. Our work shows that (contrary to expectations) close to depinning frequencies the thermal bath energy may inhibit the avalanche processes due to a nonlinear dependence of the vortex core on the microwave field. 
      
\section{Conclusions}
In \textit{Conclusion} we have systematically studied the microwave driven superconducting vortex avalanches at different frequencies and intensities of $mw$ radiation. Both the cases of weak and strong overlap of the vortex cores are investigated. The main finding is
that magnetic avalanches in microwave driven superconducting vortex systems close to depinning frequencies can become more robust to external stimulus.  This is in contrast with the usually observed thermal weakening of the avalanche process. A simple model which considers shrinking in size of the microwave driven vortex qualitatively explains the main observations. \cred{The observed effects point out towards potential robustness of superconducting devices to stimulated avalanche processes when external drive frequencies are tuned to vortex depinning frequencies.}

\section{Acknowledgments}
The authors gratefully acknowledge A. Awad and A. Silhanek for experimental help on the initial stages and K. Ilin for discussions. This work has been supported in parts by Spanish MINECO (MAT2015-66000-P), and Comunidad de Madrid (NANOFRONTMAG-CM S2013/MIT-2850) and NANO-SC COST-Action MP-1201. V. V. M. acknowledges the Methusalem Funding of the Flemish Government.

%\bibliographystyle{apsrev4-1} % Tell bibtex which bibliography style to use
%\bibliography{references}

\end{document}